\documentclass{webofc}
\usepackage[varg]{txfonts}   
\usepackage{hyperref}   
\usepackage{subcaption}
\begin{document}
\title{Comparison of bulk properties of wet granular materials using different capillary force approximations}
%
%

\author{\firstname{Meysam} \lastname{Bagheri}\inst{1} \and
        \firstname{Sudeshna} \lastname{Roy}\inst{1}\fnsep\thanks{\email{sudeshna.roy@fau.de}}\and
        \firstname{Thorsten} \lastname{P\"{o}schel}\inst{1}
}

\institute{Institute for Multiscale Simulation, \\
        Friedrich-Alexander-Universit\"at Erlangen-N\"urnberg, Erlangen, Germany 
          }

\abstract{We perform Discrete Element Method simulations of wet granular matter in a split-bottom shear cell. To calculate the capillary forces from the liquid bridges between the grains, we used three different approximations. The simulations of the shear cell showed a linear increase in bulk cohesion with the surface tension of the liquid, consistently for all approximations. However, the macroscopic friction coefficient shows only a weak dependence on surface tension.}
\maketitle
\section{Introduction}
Precise simulation of capillary forces in granular materials is essential for accurately predicting the behavior of wet granular systems. Previous studies like those of Willett \cite{willett2000capillary}, Rabinovich \cite{rabinovich2005capillary}, and Weigert \cite{weigert1999calculation} have provided valuable approximations for capillary forces, but these methods are less accurate and not easily scalable. The Willett approximation has the advantage of being relatively simple and computationally efficient, making it well-suited for large-scale simulations \cite{roy2017general,roy2016micro,gladkyy2014comparison}. It provides a reasonably accurate approximation of capillary forces for symmetric liquid bridges, particularly with small volume of liquid bridges and moderate contact angles. However, it is less accurate for larger liquid volumes, where more advanced approximation like those of Rabinovich or recent developments may offer improved precision.

Recent research has focused on improving the accuracy of these approximations for larger volumes of liquid bridges. Among these, Bagheri et al. proposed a new approximation that is particularly suitable for simulations involving polydisperse particles \cite{bagheri2024approximate}. Both the Willett \cite{willett2000capillary} and Bagheri \cite{bagheri2024approximate} approximations have been implemented in the open-source Discrete Element Method (DEM) framework, MercuryDPM \cite{bagheri2024discrete}. This work focused on comparing these approximations based on the force-interparticle distance relationship \cite{bagheri2024discrete}. The findings from these comparisons show a strong agreement between the forces predicted by the Willett and Bagheri approximations, even for particles of different sizes. 

The study by Roy et al. \cite{roy2016micro} provides a foundation for understanding the bulk properties of various capillary force approximations. Their research shows that bulk properties, like steady-state bulk cohesion, mainly depend on two factors: (i) the maximum force and (ii) the total adhesive energy dissipated per contact. Their findings suggest that if these two parameters are matched, the bulk properties will stay consistent across different capillary force approximations. Building on this foundation, our study aims to check if the bulk properties of the Willett and Bagheri approximations are truly equivalent, focusing on steady-state cohesion and the macroscopic friction coefficient. By conducting this comprehensive analysis, we seek to determine if the equivalence in force-distance relationships translates to similar bulk behaviors in practical applications.

\section{Simulation setup}
\begin{figure}[htb!]
  \begin{center}
\includegraphics[width=0.6\columnwidth]{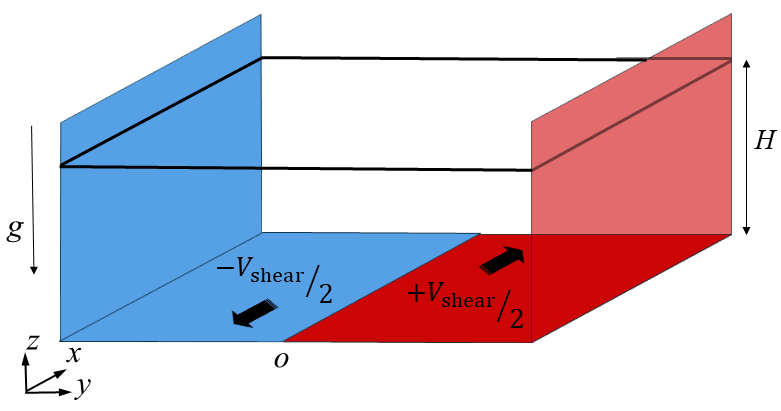}
  \end{center}
\caption{Linear split-bottom shear cell.}
\label{fig:Geometry}
 \end{figure}

We consider a numerical device to study shear bands, the linear split-bottom shear cell, which consists of two straight `L'-shaped walls sliding past each other as shown in \autoref{fig:Geometry}. The left and right `L'-shapes move along the $x$ direction in opposite directions with speed $-V_\mathrm{shear}/2$ and $V_\mathrm{shear}/2$, respectively, where $V_\mathrm{shear} = 0.016$ m/s. They are separated by a split that passes through the origin $O$. The gravitational acceleration $g$ acts in the negative $z$-direction. The particle bed consists of particles of uniform diameter $d_{\mathrm p}$. The width of the shear cell and the height of the particle bed are denoted $L$ and $H$, respectively.  We use Cartesian coordinates, where the $x$-direction is parallel and the $y$-direction is perpendicular to the split, and the $z$-direction is perpendicular to the bottom plates \cite{depken2006continuum,ries2007shear,depken2007stresses}. 
 The granular material is confined by gravity between the two `L'-shaped walls. As the two walls slide past each other in opposite directions, a shear band is formed at the center split location. The shear band is narrow at the bottom near the split location and widens at the top near the free surface. 
\section{Contact model and material parameters}
 \subsection{Liquid bridge contact model}

The nonlinear capillary force depends on particle size, contact properties, liquid properties, and the liquid saturation level in the system. Specifically, it relies on three key parameters: surface tension ($\sigma$), which determines the maximum adhesive force; contact angle ($\theta$); and liquid bridge volume ($V$), which also defines the maximum interaction distance between particles at the point of bridge rupture (rupture distance). 

The adhesive capillary force ($F_c$) between particles $i$ and $j$ is modeled using the approximations proposed by Willett et al. \cite{willett2000capillary} and Bagheri et al. \cite{bagheri2024approximate}. The Willett approximations are divided into two polynomial fits, both derived by solving the Laplace-Young equations numerically. The simplified Willett approximation is applicable for smaller liquid bridge volumes, while the classical Willett approximation applies to larger volumes and can accommodate contact angles up to $50^{\circ}$. 

\subsection{Material parameters}

All relevant material parameters are given in \autoref{tab:material_parameters}.
\begin{table}[htb!]
\caption{DEM simulation parameters.}
\label{tab:material_parameters}
\begin{center}
\begin{tabular}{l@{\quad}l@{\quad}ll}
\hline
\multicolumn{1}{l}{\rule{0pt}{18pt}
                   Variable}&\multicolumn{1}{l}{Unit}&{Value}\\[2pt]
                   \hline\rule{0pt}{12pt}\noindent
                   Elastic modulus ($E$)  & MPa& $5$\\
                   sliding friction coeff. ($\mu_s$)  & -& $0.10$\\
                   rolling friction coeff. ($\mu_r$)  & -& $0.005$\\
                   particle diameter ($d_p$)  &mm& $1.70$\\
                   Contact angle ($\theta$)  & degree& $20$\\
                   Liquid bridge volume ($V$)  & nl& $20$\\[2pt]                   
                   Particle density ($\rho$)  & kg/m$^3$& $2000$\\
                   Surface tension ($\sigma$)  & N/m& $0-0.140$\\[2pt]
                   \hline                      
\end{tabular}
\end{center}
\end{table}

\section{Micro to macro}
To extract the macroscopic fields, we employed a coarse-graining technique that calculates macro-parameters with high precision by determining sphere overlap volumes and mesh elements, as detailed by Strobl et al. \cite{strobl2016exact}. We simulated the system for 200\,s real-time. To obtain steady-state data, we average over the $x$-direction and over time from 170 to 200\,s, producing continuum fields denoted as $Q(y, z)$. We focus on the regime where the material deforms continuously at a constant strain rate without further stress changes, typically observed inside the shear band. The shear band region is defined as the area where the local strain rate at height $z$ exceeds the threshold strain rate, $\dot{\gamma}_c{(z)} \equiv 0.8\dot{\gamma}_{\text{max}}(z)$, with $\dot{\gamma}_{\text{max}}(z)$ representing the maximum strain rate at height $z$.
\subsection{Bulk cohesion from different capillary force approximations}
\begin{figure*}[htbp!]
\centering
\begin{subfigure}{0.48\linewidth}
    \includegraphics[trim={0cm 0cm 0cm 0cm},clip,width=\linewidth]{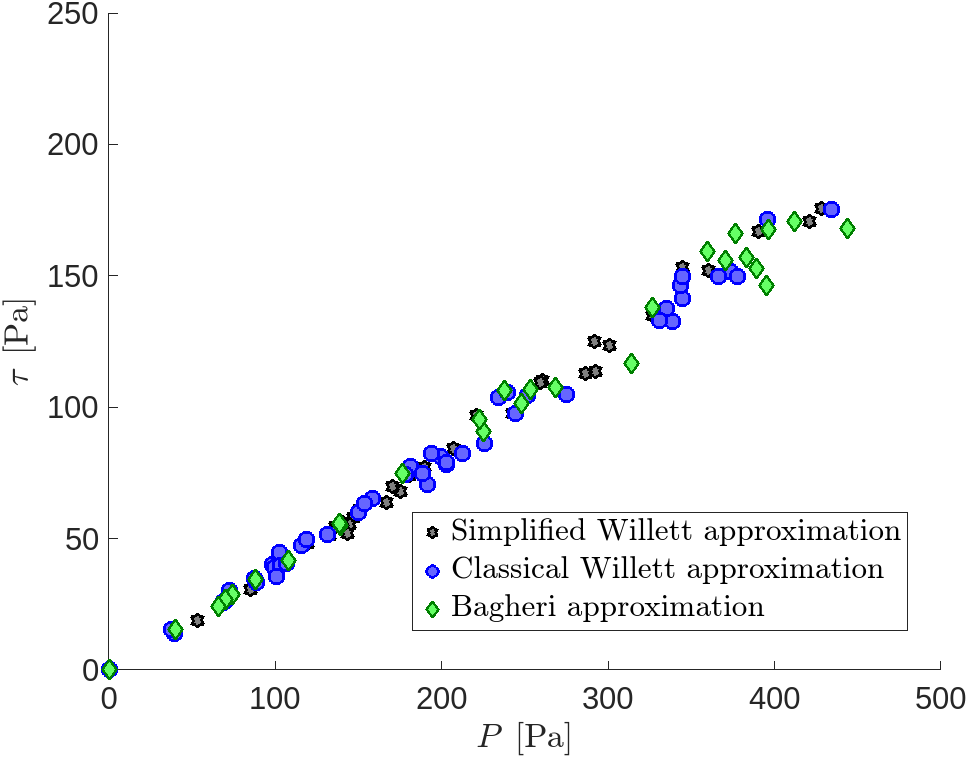}
    \subcaption{}
\end{subfigure}
\hfill
\begin{subfigure}{0.48\linewidth}
    \includegraphics[trim={0cm 0cm 0cm 0cm},clip,width=\linewidth]{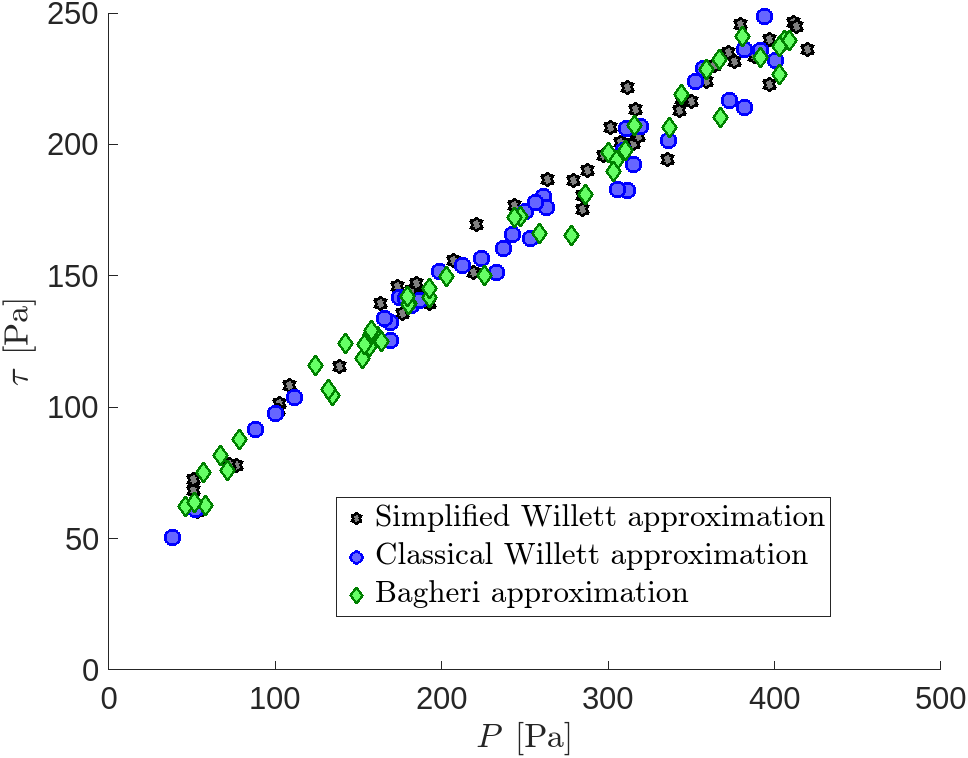}
    \subcaption{}
\end{subfigure}
\caption{Shear stress $\tau$ plotted against normal stress $P$ for wet granular materials simulated using different capillary force approximations for liquid surface tension (a) $\sigma = 0$ N/m and (b) $\sigma = 0.140$ N/m.}
\label{fig:muP} 
\end{figure*}

The local shear stress $\tau(y,z)$ plotted against the local normal stress $P(y,z)$ for data inside the shear band as shown in \autoref{fig:muP}(a) and (b) for different surface tension values of the liquid, $\sigma = 0$ N/m and $0.140$ N/m, respectively.
A linear trend is observed for the local shear stress as a function of the normal stress, which is fitted well by:
\begin{equation}\label{eq:shearstress-fit}
    \tau = \mu P+c\,,
\end{equation}
where $\mu$ is the macroscopic friction coefficient and $c$ is the steady-state cohesion obtained from linear fitting of the local shear stress and normal stress values. Note that this linear correlation is limited to intermediate to high-pressure data ($P\geq75$ Pa), with data near the free surface of the shear cell excluded from the fitting. Similar analyses on stress correlations have been established in previous studies \cite{luding2007effect, roy2016micro}. The macroscopic data $\tau$ vs. $P$ obtained from simulations using different microscopic capillary force approximations, namely the simplified Willett approximation, the classical Willett approximation and the Bagheri approximation, coincide well, as observed in the figures.

\subsection{Steady-state cohesion and its correlation with surface tension of liquid}
\begin{figure*}[htbp!]
\centering
\begin{subfigure}{0.48\linewidth}
    \includegraphics[trim={0cm 0cm 0cm 0cm},clip,width=\linewidth]{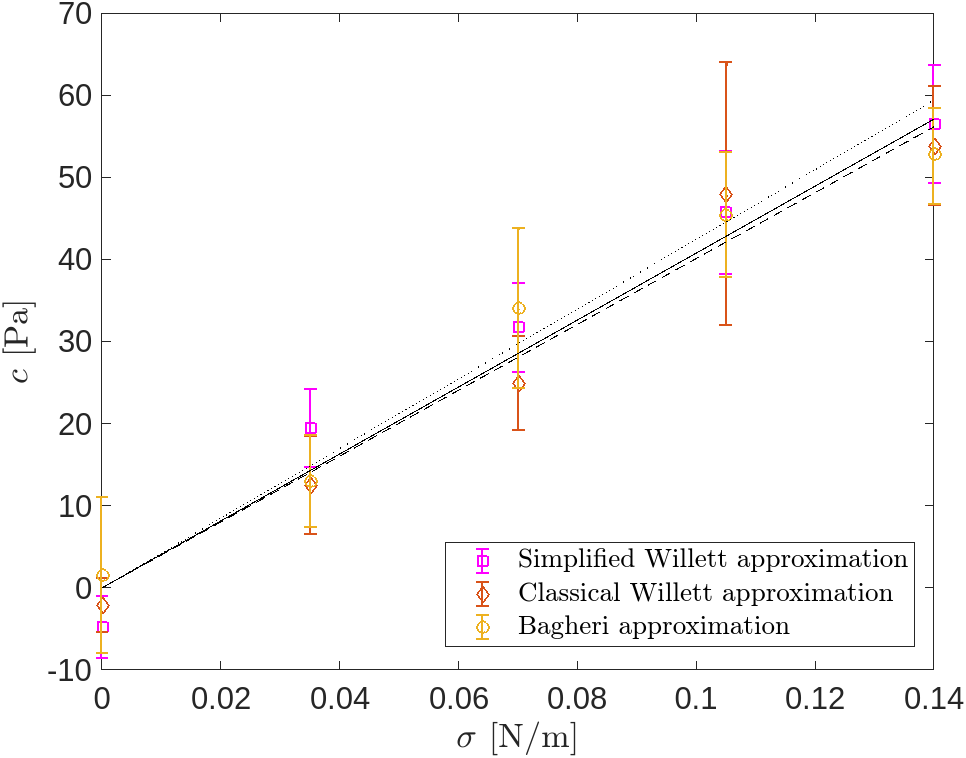}
    \subcaption{}
\end{subfigure}
\hfill
\begin{subfigure}{0.48\linewidth}
    \includegraphics[trim={0cm 0cm 0cm 0cm},clip,width=\linewidth]{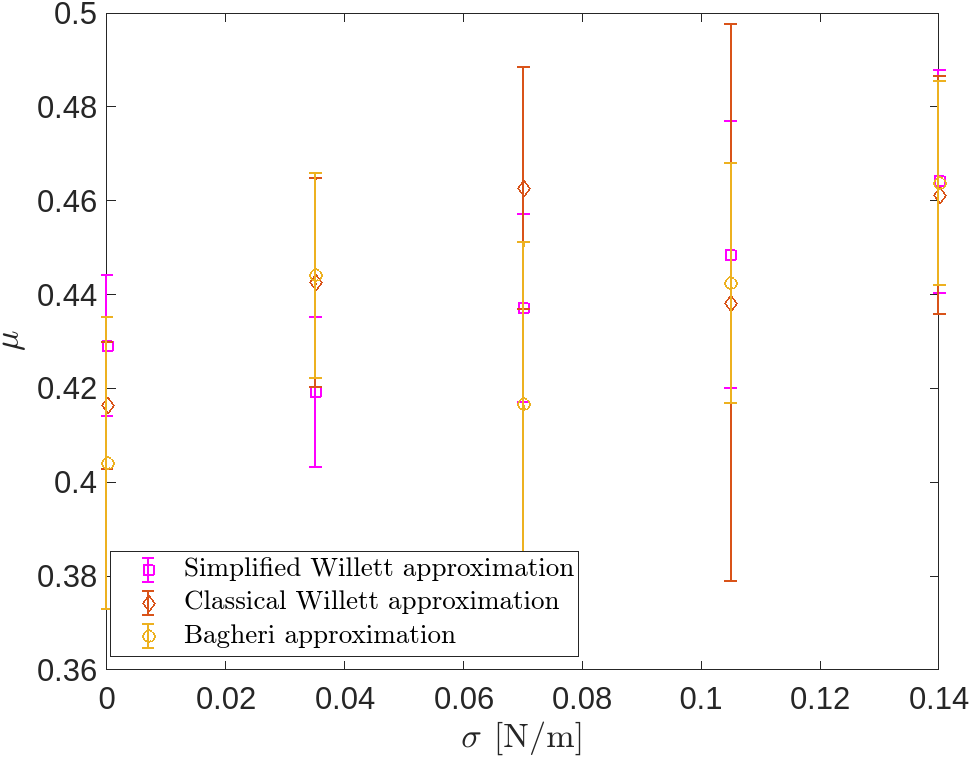}
    \subcaption{}
\end{subfigure}
\caption{(a) Bulk cohesion $c$ as a function of the surface tension of liquid $\sigma$ for liquid bridge volume $20$ nl. (b) Macroscopic friction coefficient $\mu$ as a function of the surface tension of liquid $\sigma$ for liquid bridge volume $20$ nl. The dotted, dashed and solid lines in (a) represent the linear fits to the data based on the simplified Willett, the classical Willett, and the Bagheri approximations, respectively.}
\label{fig:muC} 
\end{figure*}

We measure the steady-state cohesion $c$ and the macroscopic friction coefficient $\mu$ as functions of the liquid surface tension $\sigma$. Our previous studies show that steady-state cohesion increases linearly with increasing liquid surface tension \cite{roy2016micro}. In this study, we further examine $c$ and $\mu$ over a broader range of $\sigma \in [0, 0.140] , \text{N/m}$. \autoref{fig:muC}(a) presents the dependence of bulk cohesion $c$ on $\sigma$ for both the Bagheri and Willett approximations. The bulk cohesion for both approximations increases linearly with $\sigma$, and the linear fits, represented by the solid and dashed lines, show close agreement.
In contrast to the linear dependence of $c$ on $\sigma$, the macroscopic friction coefficient $\mu$ is very weakly dependent on $\sigma$ for both the Bagheri and Willett approximations, as shown in \autoref{fig:muC}(b). Both observations regarding the dependence of $c$ and $\mu$ on $\sigma$ are in close agreement with the findings of Roy et al. \cite{roy2016micro}.
\section{Conclusion}
We simulated wet granular materials using the simplified Willett, the classical Willett and the Bagheri approximations of capillary forces in a linear split-bottom shear cell. All the approximations showed good agreement in predicting bulk cohesion and macroscopic friction. While bulk cohesion exhibited a linear dependence on the liquid's surface tension, the macroscopic friction coefficient is very weakly dependent on the surface tension of liquid. These results confirm the applicability of different capillary force approximations and the consistency of bulk-scale properties observed in the simulations.
%
\bibliography{LiquidBridge}
%
%
%
%

\end{document}